\documentclass[aps,prl,a4paper,superscriptaddress,twocolumn,floatfix,showpacs]{revtex4}
\usepackage{graphicx}
\usepackage{latexsym}
\usepackage[]{amsmath}

\begin{document}
\newcommand{\chem}[1]{\ensuremath{\mathrm{#1}}}

\title{Partial magnetization plateau emerging from a quantum spin ice state in Tb$_2$Ti$_2$O$_7$}

\author{P.\ J.\ Baker}
\affiliation{ISIS Facility, STFC Rutherford Appleton Laboratory, Didcot OX11 0QX, United Kingdom}

\author{M.\ J.\ Matthews}
\affiliation{Department of Physics and Materials Research Institute, Pennsylvania State University, University Park, Pennsylvania 16802, USA}

\author{S.\ R.\ Giblin}
\affiliation{ISIS Facility, STFC Rutherford Appleton Laboratory, Didcot OX11 0QX, United Kingdom}

\author{P.\ Schiffer}
\affiliation{Department of Physics and Materials Research Institute, Pennsylvania State University, University Park, Pennsylvania 16802, USA}

\author{C.\ Baines}
\affiliation{Laboratory for Muon-Spin Spectroscopy, Paul Scherrer Institute, Villigen CH-5232, Switzerland}

\author{D.\ Prabhakaran}
\affiliation{Oxford University Department of Physics, Clarendon Laboratory,
Parks Road, Oxford OX1 3PU, United Kingdom}

\date{\today}

\begin{abstract}
The quantum spin ice model applied to Tb$_2$Ti$_2$O$_7$ predicts that magnetic fields applied along the $[111]$ axis will induce a partial magnetization plateau [H.\ R.\ Molavian and M.\ J.\ P.\ Gingras, J.\ Phys.: Condens.\ Matter {\bf 21}, 172201 (2009)]. We test this hypothesis using ac magnetic susceptibility and muon-spin relaxation measurements, finding features at $15$ and $65$~mT agreeing with the predicted boundaries of the magnetization plateau.
This suggests that Tb$_2$Ti$_2$O$_7$ is well described by a quantum spin ice model with an effective exchange constant of $J_{\rm eff} =0.17(1)$~K. 
\end{abstract}


\maketitle

Geometrically frustrated magnets are materials in which geometrically-induced competition between interactions prevents local magnetic moments from ordering down to temperatures well below the energy scale of those interactions.  If dynamic magnetic fluctuations persist to the lowest experimentally accessible temperatures, the low temperature magnetic state is considered a cooperative paramagnet or spin liquid.
These systems provide considerable opportunities for experimental tests of theoretical approaches to exotic collective phenomena~\cite{balents}.
In this context the pyrochlore magnet Tb$_2$Ti$_2$O$_7$ has proved an intriguing conundrum since no magnetic ordering is observed down to the lowest measured temperature, $\sim~50$~mK, far below the Curie-Weiss temperature, $\Theta_{\rm CW} = -19$~K~\cite{rmpggg}, and on this basis it has been identified as a potential three-dimensional spin liquid~\cite{balents,rmpggg,gardner99,gingras00,gardner01,yasui02,gardner03}. 
Early theoretical models describing Tb$_2$Ti$_2$O$_7$ as an antiferromagnetic $\langle 111 \rangle$ Ising system predicted long-range magnetic ordering at around $1$~K~\cite{gingras00,denhertog00}, in contradiction with the experimental results.
One route to overcoming this theoretical impasse is to consider the effects of the quantum fluctuations of the Tb$^{3+}$ magnetic moments, which renormalize the low-energy effective Hamiltonian of the system from an unfrustrated $\langle 111 \rangle$ Ising antiferromagnet to a frustrated $\langle 111 \rangle$ Ising ferromagnet~\cite{molavian07}, the quantum analog of the spin ice systems Dy$_2$Ti$_2$O$_7$ and Ho$_2$Ti$_2$O$_7$~\cite{harris,bandg}.
For Tb$_2$Ti$_2$O$_7$, this quantum spin ice model predicts a partial magnetization plateau for a magnetic field $\lesssim 0.1$~T applied along the $[111]$ crystal axis evident at low-temperatures $\lesssim 0.1$~K~\cite{molavian09}.

Perpendicular to the $[111]$ axis of a pyrochlore system lie alternating kagome and triangular layers. Applying a magnetic field along this axis in the spin ice compounds Dy$_2$Ti$_2$O$_7$ and Ho$_2$Ti$_2$O$_7$ induces a kagome ice state with an associated partial magnetization plateau that has been investigated in detail~\cite{matsuhira,moessnersondhi,tabata,fennell07}. This is analogous to the situation predicted for Tb$_2$Ti$_2$O$_7$, although the energy scales are significantly different due to the effect of the quantum fluctuations.

Perturbations are known to drive Tb$_2$Ti$_2$O$_7$ into a magnetically ordered state. Neutron scattering measurements using fields applied along the $[1\bar{1}1]$ and $[110]$ axes~\cite{yasui01,rule06,cao08,ruffxx,sazonovxx} 
and applied pressure~\cite{mirebeau00} found magnetic Bragg peaks above $0.125$~T and $2$~GPa, respectively. 
The magnetic fields used in these studies were, however, too large to test the subsequent prediction of a partial magnetization plateau. 
Measurements applying both pressure and magnetic field show that the induced magnetic ordering can be tuned using both parameters~\cite{mirebeau04}. 
Magnetoelastic effects also play their part in the low-temperature magnetic state of Tb$_2$Ti$_2$O$_7$. A giant magnetostriction $\vert \Delta l / l \vert \sim 10^{-4}$ is observed at $4.2$~K and can be understood in terms of the crystal field levels of the Tb$^{3+}$ ions~\cite{aleksandrov85}. X-ray diffraction studies point to structural fluctuations in zero magnetic field~\cite{ruff07} and find a cubic-tetragonal structural phase transition that can be resolved in magnetic fields $B \gtrsim 25$~T~\cite{ruff10}. Inelastic neutron scattering measurements made in magnetic fields applied along the $[110]$ direction show a feature at $0.04$~THz indicating the presence of a tetragonal lattice distortion~\cite{rule09}.

To test for the $[111]$ magnetization plateau in Tb$_2$Ti$_2$O$_7$ we have carried out ac susceptibility and muon spin relaxation ($\mu$SR) experiments using magnetic fields applied along the Tb$_2$Ti$_2$O$_7$ $[111]$ crystal axis. 
These provide complementary information on the magnetic susceptibility at two distinct timescales.
%
%

Our ac susceptibility measurements were performed using a custom made coil set, thermally anchored to the mixing chamber of an Oxford Instruments dilution refrigerator through immersion in liquid $^4$He.
For our $\mu$SR experiments~\cite{blundell99} the field was parallel to the initial muon spin polarization [longitudinal field (LF)] in the ISIS measurements ($0.04 < T < 10$~K) and with the initial muon spin polarization partially rotated [LF and transverse field (TF)] for measurements ($0.025 < T < 0.9$~K) using the Low Temperature Fridge (LTF) spectrometer (Paul Scherrer Institute, Switzerland). 
The measured parameter is the time-dependent muon decay asymmetry, $A(t)$, recorded in positron detectors on opposite sides of the sample. Having subtracted the background and normalized the signal, this provides a measure of the spin polarization $P(t) = [A(t)-A_{\rm bg}]/[A(0)-A_{\rm bg}]$ of the muon ensemble as a function of time. For all our $\mu$SR experiments the Tb$_2$Ti$_2$O$_7$ crystals, grown using a floating zone furnace, were arranged in a mosaic and attached to a silver backing plate using a thin layer of GE varnish. The silver plate gives a temperature and field-independent background signal that can easily be identified and subtracted from the asymmetry data.

\begin{figure}[t]
\includegraphics[width=\columnwidth]{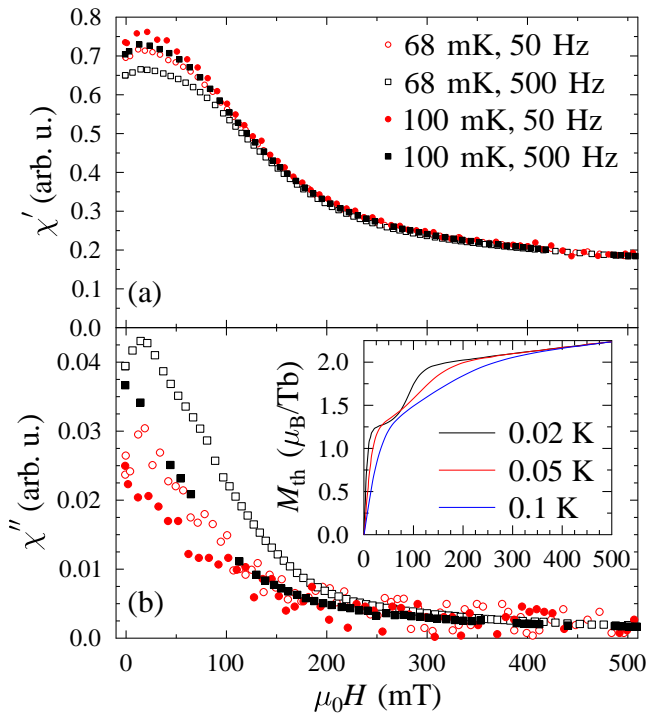}
\caption{ 
(Color online) ac susceptibility data with the field applied $\parallel [111]$:
(a) Real part $\chi^{\prime}$,
(b) Imaginary part $\chi^{\prime\prime}$.
(Inset) Theoretical magnetization curves predicted by the quantum spin ice model of Ref.~\onlinecite{molavian09} for $J=0.167$~K.
\label{acdata}}
\end{figure}

The ac susceptibility data are shown in Figure~\ref{acdata}~(a) for the real part $\chi^{\prime}$ and (b) for the imaginary part $\chi^{\prime\prime}$. The data recorded at $100$ and $125$~mK are equivalent within error ($125$~mK data not shown) but there is a clear separation between these data and those recorded at $68$~mK in both components of the susceptibility. In $\chi^{\prime}$ the separation is clear below $\sim 200$~mT and grows towards zero field. This can be compared to the theoretically predicted magnetization curves~\cite{molavian09} in the inset to Fig.~\ref{acdata}, where a separation between the predictions for $50$ and $100$~mK emerges below $\sim 300$~mT. The temperature dependence is more pronounced in $\chi^{\prime\prime}$, where at $68$~mK the susceptibility rises to a peak at $15$~mT but at $\geq 100$~mK it falls monotonically. As with $\chi^{\prime}$, at $68$~mK $\chi^{\prime\prime}$ remains distinct from that at $100$~mK up to $\sim 300$~mT, in accordance with the behavior predicted for the magnetization. Any feature at the upper boundary of the magnetization plateau is indistinct, also as predicted. 
Previous measurements of the ac susceptibility indicated a partial spin freezing at low temperature and a frequency dependence consistent with our results~\cite{gardner03,hamaguchi04}.

\begin{figure}[t]
\includegraphics[width=\columnwidth]{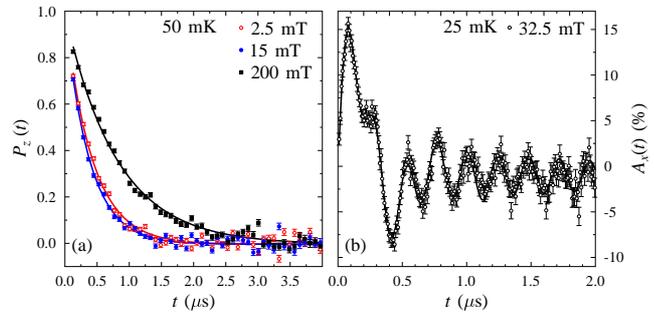}
\caption{ 
(Color online) 
(a) Longitudinal field $\mu$SR data at $50$~mK.
(b) Transverse field $\mu$SR data at $25$~mK. The phase shift is due to the orientation of the positron detectors with respect to the initial muon polarization.
The solid lines are the fits to the data described in the text.
\label{musrdata}}
\end{figure}

The $\mu$SR data recorded in longitudinal field are shown in Fig.~\ref{musrdata}~(a) and can all be described by the equation:
\begin{equation}
P_{z}(t) = e^{-\lambda t},
\label{LFrawfit}
\end{equation}
where $\lambda$ is the muon spin relaxation rate. The background is weakly relaxing in zero applied field but effectively constant in longitudinal applied field, consistent with the expected behaviour of the silver sample holder. That this equation describes the data over the whole measured field range demonstrates that the magnetic fields remain dynamic on the timescale probed by muons.
The transverse field data shown in Fig.~\ref{musrdata}~(b) take a similar form except for the muon spin precession. They can be described by the equation:
\begin{equation}
A_{x}(t) = A_{\rm s}e^{-\lambda_{\rm s} t}\sin(\omega_{\rm s} t) + A_{\rm bg}e^{-\lambda_{\rm bg} t}\sin(\omega_{\rm bg} t),
\label{TFrawfit}
\end{equation}
where the first term describes the signal from the sample and the second term describes the background signal. The oscillations are sinusoidal rather than the conventional cosinusoidal behavior because of the detector geometry used and the angular precession frequencies are related to the magnetic fields experienced by the muons as $\omega = \gamma_{\mu}B$ ($\gamma_{\mu}/2\pi = 135.5$~MHzT$^{-1}$). 

The magnetic fluctuations in Tb$_2$Ti$_2$O$_7$ have previously been shown to be fast compared with the range of timescales probed by muons~\cite{gardner99,dunsiger,keren04,ofer07}, that is to say that the system is paramagnetic on the muon timescale, and the relaxation rate $\lambda$ can therefore be related to the distribution width of magnetic fields at the muon stopping site $\Delta$, the fluctuation time $\tau$, and the applied longitudinal field $B_{\rm LF}$ by the sum of Redfield's equation~\cite{slichter} and a field-independent relaxation rate, $\lambda_0$:
\begin{equation}
\lambda = \frac{2\gamma^{2}_{\mu}\Delta^{2}\tau}{1+\gamma^{2}_{\mu}B^{2}_{\rm LF}\tau^2} + \lambda_{0}.
\label{redfield}
\end{equation} 
We include the field-independent relaxation rate, $\lambda_0$, because our data appear to tend to a field-independent value above $0.25$~T, which is consistent with the behaviour seen in that field range in Ref.~\cite{dunsiger}. 

\begin{figure}[t]
\includegraphics[width=8.5cm]{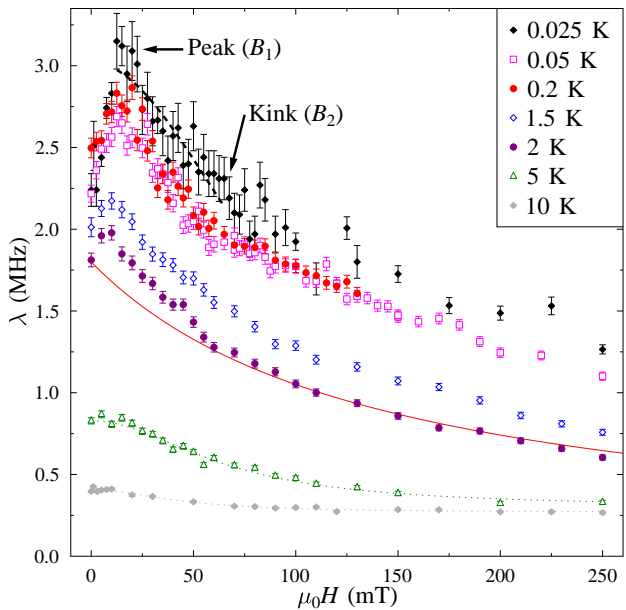}
\caption{
(Color online)
Relaxation rate $\lambda$ measured as a function of longitudinal magnetic field applied along the $[111]$ axis. Data recorded at $0.5$ and $0.725$~K are similar to those shown for $T \leq 0.2$~K so are omitted for clarity. The solid line represents the trend for a polycrystalline sample at $0.1$~K reported in Ref.~\onlinecite{keren04}. The dashed and dotted lines represent fits described in the text.
\label{LFdata}}
\end{figure}

For the $T \leq 2$~K longitudinal field data, three regions can be identified in the field dependent relaxation rates shown in Fig.~\ref{LFdata}.
At small fields up to $B_1 \sim 15$~mT, $\lambda$ increases to a peak, then falls steeply to a kink at $B_2 \sim 60$~mT, followed by a more gradual fall between $60$ and $250$~mT. The fields at which the peak and kink are observed are consistent with the boundaries of the magnetization plateau predicted in Ref.~\cite{molavian09} (see the inset to Fig.~\ref{acdata}). For comparison, we plot the trend $\lambda(B_{\rm LF}) \propto B^{-1}$ previously found for polycrystalline data at $100$~mK~\cite{keren04} as a solid line in Fig.~\ref{LFdata}. This is similar to the behaviour seen in our data above $\sim 70$~mT.

Using equation~\ref{redfield}, we can test predictions for the field dependence of $\Delta$ and $\tau$. The simplest assumption is that neither depends on field, which effectively describes both the $5$ and $10$~K data. Below $5$~K this model does not work over the whole field range. However, it is effective between the peak ($B_1$) and kink ($B_2$) shown in Fig.~\ref{LFdata}, thereby suggesting that a plateau exists in the local magnetic field distribution in the anticipated field range. Fitting (with $\lambda_0 = 0$) leads to $\Delta \sim 12$~mT and $\tau \sim 13$~ns at low-temperature, the fit for $25$~mK being shown as the dashed line in Figure~\ref{LFdata}. Including the $\lambda_0$ values estimated using the data above $150$~mT ($\sim 0.5$~MHz) reduces the value of $\Delta$ by around $15$~\% and increases $\tau$ by around $10$~\%. Independent of this, the quality of the fits is poorer for the $1.5$ and $2$~K data, which is consistent with thermal fluctuations breaking up the low temperature state. Above $\sim 65$~mT it is not possible to describe the data using equation~\ref{redfield} and the same parameters as between $15$ and $60$~mT. This strongly suggests that either the field distribution or the fluctuation timescale has changed.

\begin{figure}[t]
\includegraphics[width=\columnwidth]{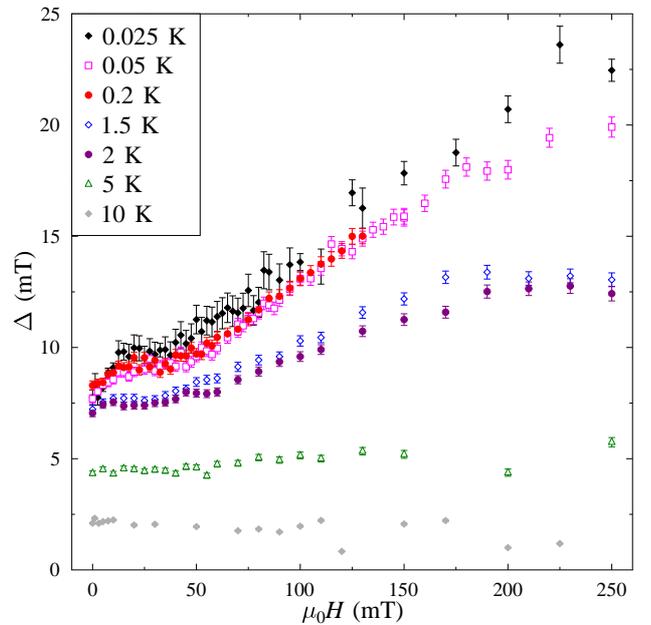}
\caption{
(Color online) 
Field dependence of the local field distribution $\Delta$ estimated from the $\lambda$ values in Fig.~\ref{LFdata} using equation~\ref{Delta}, $\tau = 20$~ns, and the $\lambda_0$ values determined from fits to the high-field data. 
\label{LFparameters}}
\end{figure}

To visualize the field-dependence of the local magnetic field distribution $\Delta$ under the assumption that the fluctuation timescale $\tau$ is independent of the weak magnetic fields being applied along $[111]$, we can rearrange equation~\ref{redfield} into the form:
\begin{equation}
\Delta = \sqrt{(\lambda-\lambda_0)(1+\gamma^{2}_{\mu}B^{2}\tau^{2})/2\gamma^{2}_{\mu}\tau}.
\label{Delta}
\end{equation}
Fig.~\ref{LFparameters} shows the resulting $\Delta$ values assuming $\tau = 20$~ns at all temperatures. This $\tau$ value is suggested by the fits to the higher temperature data although the form of these results is only weakly dependent on $\tau$ for $12 < \tau < 25$~ns, with the plateau region consistently evident in the $T \leq 2$~K data. We expect the fits to equation~\ref{redfield} in the plateau region will underestimate $\tau$ since $\Delta$ is assumed constant, which is unlikely at non-zero temperature. 
The $\lambda_0$ values for $T \leq 2$~K were estimated by fitting equation~\ref{redfield} with constant $\Delta$ and $\tau$ to the highest field data available, $\mu_{0}H \gtrsim 150$~mT, and seeking a value that did not change significantly with the low-field boundary of the fitting window. 
In the inset to Fig.~\ref{acdata} we show the theoretical magnetization curves from Ref.~\onlinecite{molavian09}. There is good agreement in the position of the plateau region. Since we measure the local magnetic field distribution width $\Delta$ rather than $M$, it is not possible to make quantitative comparisons of the magnitude. 

We also carried out muon-spin rotation measurements from $2.5$ to $250$~mT at $25$~mK with the aim of finding evidence for the level crossing predicted to occur at the upper boundary of the magnetization plateau by the theory described in Ref.~\cite{molavian09}. No evidence for such an effect was found, which could be due to correlations between neighbouring tetrahedra or, alternatively, that the primary relaxation mechanism for the muon spin is always the large distribution of fluctuating local fields. Instead we observed a significant negative frequency shift between the precession frequencies in the sample and the background, $K = (\omega_{\rm s} - \omega_{\rm bg})/\omega_{\rm bg}$. Above $\sim 7.5$~mT, $-0.65 < K < -0.625$ with negligible field dependence. This value is consistent with previous measurements at higher fields~\cite{dunsiger,ofer07}, which attributed the effect to the large sample magnetization.

In conclusion, both ac susceptibility and $\mu$SR measurements of the dynamic magnetic fields in Tb$_2$Ti$_2$O$_7$ show features suggesting that a partial magnetization plateau emerges for fields between $12.5$ and $65$~mT applied along the $[111]$ axis. Comparing these fields to the predictions of the quantum spin ice model suggests an effective exchange constant of $J_{\rm eff} = 0.17(1)$~K. The two probes show the features being smeared away by thermal fluctuations at distinct temperatures - for ac susceptibility this is around $100$~mK, similar to that predicted for the bulk magnetization~\cite{molavian09}, whereas for $\mu$SR the features persist to around $2$~K, consistent with the temperature dependence of the relaxation rate observed at a constant field in previous measurements~\cite{gardner99}. The higher temperature scale associated with the $\mu$SR results implies that on shorter timescales, estimated from fitting the field dependence in the plateau region to be $\tau \simeq 20$~ns, the fluctuations persist to higher temperature. Such fluctuations would be quasistatic on the neutron timescale and therefore extending previous neutron scattering studies to lower fields would be worthwhile, together with bulk measurements of the magnetization. Very recently an alternative theoretical approach to describing the observed properties of Tb$_2$Ti$_2$O$_7$ was proposed, based on a Jahn-Teller-like distortion and a two-singlet system coupled by exchange~\cite{bonvillexx}. Our results provide a further quantitative test for this model.

Parts of this work were performed at the ISIS Facility, UK, and at the Swiss Muon Source, Paul Scherrer Institute, Villigen, CH. We thank Andrew Boothroyd for help with crystal preparation, Larry Linfitt and the ISIS cryogenics group for experimental assistance, Francis Pratt and Michel Gingras for helpful discussions, STFC (UK) for provision of beamtime, and NSF grant DMR-0701582 for support of MJM and PS.



\begin{thebibliography}{xx}
\bibitem{balents}
L.\ Balents, Nature {\bf 464}, 199 (2010).
\bibitem{rmpggg}
J.\ S.\ Gardner, M.\ J.\ P.\ Gingras, and J.\ E.\ Greedan, Rev.\ Mod.\ Phys.\ {\bf 82}, 53 (2010).
\bibitem{gardner99} 
J. S. Gardner {\em et al.}, 
Phys. Rev. Lett. {\bf 82}, 1012 (1999).
\bibitem{gingras00}
M. J. P. Gingras {\em et al.}, Phys.\ Rev.\ B {\bf 62}, 6496 (2000).
\bibitem{gardner01}
J.\ S.\ Gardner {\em et al.},
Phys.\ Rev.\ B {\bf 64}, 224416 (2001).
\bibitem{yasui02} 
Y.\ Yasui {\em et al.},
J.\ Phys.\ Soc.\ Japan {\bf 71}, 599 (2002).
\bibitem{gardner03} 
J.\ S.\ Gardner {\em et al.}, 
Phys.\ Rev.\ B {\bf 68}, 180401(R) (2003). 
\bibitem{denhertog00}
B.\ C.\ {den Hertog} and M.\ J.\ P.\ Gingras, Phys.\ Rev.\ Lett.\ {\bf 84}, 3430 (2000).
\bibitem{molavian07}
H.\ R.\ Molavian, M.\ J.\ P.\ Gingras, and B.\ Canals, Phys.\ Rev.\ Lett.\ {\bf 98}, 157204 (2007).
\bibitem{harris}
M.\ J.\ Harris, S.\ T.\ Bramwell, D.\ F.\ McMorrow, T.\ Zeiske, and K.\ W.\ Godfrey, Phys. Rev. Lett. {\bf 79}, 2554 (1997).  
\bibitem{bandg}
S.\ T.\ Bramwell and M.\ J.\ P.\ Gingras, Science {\bf 294}, 1495 (2001).
\bibitem{molavian09}
H.\ R.\ Molavian and M.\ J.\ P.\ Gingras, J.\ Phys.: Condens.\ Matter {\bf 21}, 172201 (2009).
\bibitem{matsuhira}
K.\ Matsuhira, Z.\ Hiroi, T.\ Tayama, S.\ Takagi, and T.\ Sakakibara, J. Phys. Condens. Matter {\bf 14}, L559 (2002). 
\bibitem{moessnersondhi}
R.\ Moessner and S.\ L.\ Sondhi, Phys.\ Rev.\ B {\bf 68}, 064411 (2003).
\bibitem{tabata}
Y.\ Tabata, {\em et al.}, 
Phys. Rev. Lett. {\bf 97}, 257205 (2006). 
\bibitem{fennell07}
T. Fennell {\em et al.},
Nature Physics {\bf 3}, 566 (2007).
\bibitem{yasui01}
Y.\ Yasui, M.\ Kanada, M.\ Ito, H.\ Harashinaa, M.\ Sato, H.\ Okumura, and K.\ Kakurai, J.\ Phys.\ Chem. Solids {\bf 62}, 343 (2001).
\bibitem{rule06}
K.\ C.\ Rule {\em et al.},
Phys.\ Rev.\ Lett.\ {\bf 96}, 177201 (2006).
\bibitem{cao08}
H.\ Cao {\em et al.},
Phys.\ Rev.\ Lett.\ {\bf 101}, 196402 (2008).
\bibitem{ruffxx}
J.\ P.\ C.\ Ruff, B.\ D.\ Gaulin, K.\ C.\ Rule, and J.\ S.\ Gardner,
Phys.\ Rev.\ B {\bf 82}, 100401(R) (2010).
\bibitem{sazonovxx}
A.\ P.\ Sazonov {\em et al.}
Phys.\ Rev.\ B {\bf 82}, 174406 (2010).
\bibitem{mirebeau00}
I.\ Mirebeau, I.\ N.\ Goncharenko, P.\ Cadavez-Peres, S.\ T.\ Bramwell, M.\ J.\ P.\ Gingras, and J. S. Gardner, Nature {\bf 420}, 54 (2002).
\bibitem{mirebeau04}
I.\ Mirebeau, I.\ N.\ Goncharenko, G.\ Dhalenne, and A.\ Revcolevschi, 
Phys.\ Rev.\ Lett.\ {\bf 93}, 187204 (2004).
\bibitem{aleksandrov85}
I.\ V.\ Aleksandrov {\em et al.}, 
Sov. Phys. JETP {\bf 62}, 1287 (1985).
\bibitem{ruff07}
J.\ P.\ C.\ Ruff {\em et al.}, 
Phys.\ Rev.\ Lett.\ {\bf 99}, 237202 (2007).
\bibitem{ruff10}
J.\ P.\ C.\ Ruff {\em et al.}, 
Phys.\ Rev.\ Lett.\ {\bf 105}, 077203 (2010).
\bibitem{rule09}
K.\ C.\ Rule and P.\ Bonville, J.\ Phys.: Conf.\ Ser. {\bf 145}, 012027 (2009).
\bibitem{blundell99} 
S. J. Blundell, Contemp. Phys. {\bf 40}, 175 (1999).
\bibitem{hamaguchi04}
N.\ Hamaguchi {\em et al.},
Phys.\ Rev.\ B {\bf 69}, 132413 (2004).
\bibitem{dunsiger}
S.\ R.\ Dunsiger, Ph.D. thesis, University of British Columbia, 2000. 
\bibitem{keren04}
A.\ Keren, J.\ S.\ Gardner, G.\ Ehlers, A.\ Fukaya, E.\ Segal, and Y.\ J.\ Uemura, Phys.\ Rev.\ Lett.\ {\bf 92}, 107204 (2004). 
\bibitem{ofer07}
O.\ Ofer, A.\ Keren, and C.\ Baines, J.\ Phys.: Condens.\ Matter {\bf 19}, 145270 (2007). 
\bibitem{slichter}
C.\ P.\ Slichter, {\it Principles of Magnetic Resonance} (3$^{\rm rd}$ edition, Springer-Verlag, New York, 1996).
\bibitem{bonvillexx}
P.\ Bonville, I.\ Mirebeau, A.\ Gukasov, S.\ Petit, and J.\ Robert, arXiv:1104.1584 (unpublished).
\end{thebibliography}
\end{document}